\documentclass[showkeys,showpacs,preprintnumbers,aps]{revtex4}
\usepackage{graphicx}
\usepackage{dcolumn}
\usepackage{amssymb}
\usepackage{bm}
\usepackage[sumlimits,intlimits]{amsmath}
\usepackage{amsfonts,amssymb}
\usepackage{amsmath}
\usepackage{bm}
\usepackage{graphics}
\usepackage{fancyhdr}
\usepackage{rotating}
\DeclareGraphicsExtensions{.pdf,.jpg,.eps}
\usepackage{subfigure}
\usepackage{textcomp}
\usepackage{color}
\usepackage{rotating}
\usepackage{float}
\begin{document}
\baselineskip 16pt

\title{Time-dependent toroidal compactification proposals and the Bianchi type II model: classical and quantum solutions}
\author{J. Socorro}
\email{socorro@fisica.ugto.mx}
\author{L. Toledo Sesma,}
\email{ltoledo@fisica.ugto.mx} \affiliation{Departamento de
F\'{\i}sica, DCeI, Universidad de Guanajuato-Campus Le\'on,
C.P. 37150, Le\'on, Guanajuato, M\'exico}

\begin{abstract} In this work we construct an effective four-dimensional model by compactifying a ten-dimensional theory of
gravity coupled with a real scalar dilaton field on a time-dependent
torus without the contributions of fluxes as first approximation.
This approach is applied to anisotropic cosmological Bianchi type II
model for which we study the classical coupling of the anisotropic
scale factors with the two real scalar moduli produced by the
compactification process. Also, we present some solutions to the
corresponding Wheeler-DeWitt (WDW) equation in the context of
Standard Quantum Cosmology and we claim that these quantum solution
are generic in the moduli scalar field for all Bianchi Class A
models. Also we gives the relation to these solutions for asymptotic
behavior to large argument in the corresponding quantum solution in
the gravitational variables and is compared with the Bohm's
solutions, finding that this corresponds to lowest-order WKB
approximation.
\end{abstract}
\date{\today}
\maketitle

\section{Introduction}
In the last years there have been several attempts to understand the diverse aspects of cosmology, as the presence of stable
vacua and inflationary conditions,  in the framework of  supergravity and string theory \cite{Damour0,Horne,Banks,Damour1,
Banks1,Kallosh0,Kallosh1,Baumann}. One of the most interesting features emerging from these type of models consists on the
study of the consequences of higher dimensional degrees of freedom on the cosmology derived from four-dimensional effective
theories \cite{Khoury}.

Furthermore, it has been pointed out that the presence of extra dimensions leads to an interesting connection with the
ekpyrotic model \cite{Khoury}, which  generated considerable activity \cite{Kallosh0,Kallosh1,Enqvist}. The essential
ingredient in these models (see for instance \cite{Khoury}) is to consider an effective action with a graviton and a massless
scalar field, the dilaton, describing the evolution of the Universe, while incorporating some of the ideas of pre-big-bang
proposal \cite{Veneziano} in that the evolution of the Universe began in the far past.

On the other hand, it is well known that relativistic theories of gravity such as general relativity or string theories are
invariant under reparametrization of time. The quantization of such theories presents a number of problems of principle  known
as ``the problem of time'' \cite{Isham,Kuchar}. This problem is present in all systems whose classical version is invariant
under time reparametrization, leading to  its absence at the quantum level. Therefore, the formal question involves how to
handle the classical Hamiltonian constraint, $\mathcal{H} \approx 0$, in the quantum theory. Also, connected with the problem
of time is the ``Hilbert space problem'' \cite{Isham,Kuchar} referring to the not at all obvious selection of the  inner
product of states in quantum gravity, and whether there is a need for such a structure at all.

The above features, as it is well known, point out the necessity to construct a consistent theory of gravity at quantum level.
One promising attempt is string theory where the structure of the internal space plays a relevant role in the construction of
interesting effective models containing cosmological features as the presence of positive valued minima of the scalar potential
and slow-roll inflationary conditions. The usual procedure for that is to consider compactification on generalized manifolds,
on which internal fluxes have back-reacted, altering the smooth Calabi-Yau geometry  and stabilizing all the moduli
\cite{Baumann}.

In the present work we shall consider an alternative procedure about the role played by the moduli. In particular we shall not
consider the presence of fluxes, as in string theory, in order to obtain a moduli-dependent scalar potential in the effective
theory. Rather, we are going to promote some of the moduli to  time-dependent fields by considering  the particular case of a
ten-dimensional gravity coupled to a time-dependent dilaton compactified on a 6-dimensional torus with a time-dependent
K\"ahler modulus. With the propose  to track down the role play by such  fields, we are going to ignore the dynamics of the
complex structure field (for instance, by assuming that it is already stabilized by the presence  of a string field in
higher scales).

The work is organized in the following form. In section II we
present the construction of our effective action by compactification
on a time-dependent torus,  while in  Section III we study its
Lagrangian and Hamiltonian descriptions using as toy model the
Bianchi type II cosmological model and we present the general
structure for all Bianchi Class A models. Section IV is devoted on
finding the corresponding classical solutions for few different
cases involving the presence or absence of matter content. In
Section V we present some solutions to the corresponding
Wheeler-DeWitt (WDW) equation in the context of Standard Quantum
Cosmology, resulting the modified Bessel function $\rm (K_\mu(z),
I_\mu(z))$, solutions that in some sense are similar to this find in
previous work \cite{soc2}. In order that the unnormalized
probability density $|\Theta(\Omega,\beta_\pm)|^2$ does not diverge
for $\rm |\beta_\pm|\to \infty$ and at fixed $\Omega$, which are the
gravitational variables, we drop the function $\rm I_\mu(z)$,
remaining only the function $\rm K_\mu(z)$. Also in this section we
make an analysis for particular case in the constants in the
solution found in the asymptotic behavior for large argument in the
gravitational variables and it is compare with the Bohm's solutions
obtained, finding that this corresponds to lowest-order WKB
approximation. The Bohm's solutions was obtained in other
decomposition of the quantum WDW equation, related with the Bohm's
formalism, which was applied in the supersymmetric quantum cosmology
\cite{osb}. Finally our conclusions are presented in Section VI.

\section{Effective model}
We start from a ten-dimensional action coupled with a dilaton (which is the bosonic component common to all superstring
theories), which after dimensional reduction can be interpreted as a Brans-Dicke like theory \cite{Brans}. In the string frame,
the effective action depends on two time-dependent scalar fields: the dilation $\Phi(t)$ and the K\"ahler modulus $\sigma(t)$,
where the initial high-dimensional (effective) theory is given by

\begin{equation}
S=\frac{1}{2 \kappa^2_{10}} \int d^{10}X
\sqrt{-\hat{G}}\,e^{-2\Phi}\,\Bigg[\hat{\mathcal{R}}^{(10)}[\hat{G}_{MN}]
+ 4\,\hat{G}^{MN}\nabla_{M}\,\Phi\,\nabla_{N}\,\Phi\Bigg],
\label{action}
\end{equation}
with a metric described by

\begin{equation}
ds^2=\hat{G}_{MN}\,dX^{M}\,dX^{N}=g_{\mu\nu}\,dx^{\mu}\,dx^{\nu}+
h_{mn}\,dy^{m}\,dy^{n}, \label{metric}
\end{equation}
where $M, N, P, \ldots$ are the indices of the 10-dimensional space,
and greek and latin indices correspond to the external and internal
space, respectively.  The internal metric is given by

\begin{equation}
h_{mn}=e^{-2\sigma(t)}\delta_{mn}. \label{intmet}
\end{equation}
In order to rewrite it in the Einstein frame, we take as usual, the conformal transformation of the four dimensional metric
$g_{\mu\nu}$ as

\begin{equation}
g_{\mu\nu}= e^{2\phi(t)} g^{E}_{\mu\nu}, \label{metra}
\end{equation}
where $g_{\mu\nu}$ is the metric in the string frame and $g^{E}_{\mu\nu}$ is the metric in Einstein frame and the dilaton has
been redefined accordingly as $\phi=\Phi - \frac{1}{2}\,\ln(Vol(X_6))$. Notice that in this way, the effective dilaton
$\phi$ is also a time-dependent field. Following this notion, we have that the action \eqref{action} reads

\begin{equation}
S=\frac{1}{2\kappa^2_{4}}\int d^4x\sqrt{-g}
\bigg(\mathcal{R} - 6g^{\mu \nu}\nabla_{\mu} \nabla_\nu \phi -
6g^{\mu \nu}\nabla_{\mu} \nabla _\nu {\sigma} - 2g^{\mu \nu}
\nabla_\mu \phi \nabla_\nu \phi - 96g^{\mu
\nu}\nabla_\mu \sigma \nabla_\nu {\sigma} - 36g^{\mu \nu}\nabla_\mu\phi\nabla_\nu\sigma\bigg),
\label{redact}
\end{equation}
where we have omitted the upper script $E$ in the metric. Now with the expression \eqref{redact} we proceed to build the
Lagrangian and the Hamiltonian of the theory at the classical regimen employing the anisotropic cosmological Bianchi type II
model.

\section{The classical Hamiltonian}

In the last section we have built the reduced effective action, now we will construct the classical Hamiltonian from the
expression \eqref{redact}. We are going to assume that the background of the extended space (four dimensional) is a Bianchi
type II anisotropic cosmological model. In order to do it, let us recall here the canonical formulation in the ADM formalism
of the diagonal Bianchi Class A cosmological models. The metric has the form

\begin{equation}
ds^2= -N(t)dt^2 + e^{2\Omega(t)}\, (e^{2\beta(t)})_{ij}\,\omega^i \,
\omega^j,
\label{met}
\end{equation}
where $N(t)$ is the lapse function, $\beta_{ij}(t)$ is a $3\times3$ diagonal matrix,
$\beta_{ij}=\text{diag}(\beta_++ \sqrt{3}\beta_-,\beta_+- \sqrt{3} \beta_-, -2\beta_+)$, $\Omega(t)$ and $\beta_\pm$ are scalar
functions known as Misner variables, $\omega^i$ are one-forms that characterize each cosmological Bianchi type model
\cite{Ryan}, and obey the form $d\omega^i= \frac{1}{2} C^i_{jk} \omega^j\wedge\omega^k$, with $C^i_{jk}$ the structure
constants of the corresponding model. The one-forms for the Bianchi type II model are
$\omega^1= dx^2 - x^1dx^3$, $\omega^2= dx^3$, $\omega^3=dx^1$. So, the corresponding metric of the Bianchi type II in
Misner's parametrization has the form

\begin{align*}
ds^2=&-N^2dt^2 + e^{2\Omega-4\beta_{+}}\,\left(dx^1\right)^2 + e^{2\Omega + 2\beta_{+} + 2\sqrt{3}\beta_{-}}
\,\left(dx^2\right)^2 + \left(\left(x^1\right)^2 e^{2\Omega + 2\beta_{+} + 2\sqrt{3}\beta_{-}} +
e^{2\Omega + 2\beta_{+} - 2\sqrt{3}\beta_{-}}\right)\,\left(dx^3\right)^2\nonumber\\
&- 2x^1 e^{2\Omega + 2\beta_{+} + 2\sqrt{3}\beta_{-}}\,dx^2 dx^3.
\end{align*}
substituting $x^1=x$, $x^2=y$, $x^3=z$, we have the following form

\begin{align}
ds^2=&-N^2dt^2 + e^{2\Omega - 4\beta_{+}}\,dx^2 + e^{2\Omega + 2\beta_{+} + 2\sqrt{3}\beta_{-}}\,dy^2 +
\left(x^2\,e^{2\Omega + 2\beta_{+} + 2\sqrt{3}\beta_{-}} + e^{2\Omega + 2\beta_{+} - 2\sqrt{3}\beta_{-}}\right)\,dz^2\label{4}\\
&-2x\,e^{2\Omega + 2\beta_{+} + 2\sqrt{3}\beta_{-}}\,dy\,dz.\nonumber
\end{align}
The Lagrangian density as effective theory for this model, including
an intuitive way the matter content for a barotropic perfect fluid
as a first approximation, has the structure
\cite{Socorro,Mathur,odin1,odin2}

\begin{equation*}
\mathcal{L}_{mat}=16 \pi G_N \sqrt{-g} \rho=16N\pi G_N C_\gamma
e^{-3(1+\gamma) \Omega}
\end{equation*}
where $\gamma$ is the parameter that characterize the  epochs in the evolution to the Universe. The Lagrangian that describes
this model is given by

\begin{equation}
\mathcal{L}_{II}= \frac{e^{3\Omega}}{N}\left[ 6\,\dot{\Omega}^2 - 6\,\dot{\beta}^2_{+} - 6\,\dot{\beta}^2_{-}
+ 96\,\dot{\sigma}^2 + 36\,\dot{\phi}\,\dot{\sigma} + 2\,\dot{\phi}^2 + 16\,\pi\,G\,N^2\,\rho
+ \frac{1}{2}N^2 e^{-2\Omega + 4\beta_{+} + 4\sqrt{3}\beta_{-}}\right].
\label{lagra-i}
\end{equation}
\\*
Computing the momenta associated to the moduli fields $(\phi, \sigma)$ and Misner variables $(\Omega, \beta_{+}, \beta_{-})$
and using the Legendre transformation we obtain the Hamiltonian density for this model

\begin{equation}
\mathcal{H}_{II}=\frac{e^{-3\Omega}}{24}\left[\Pi_\Omega^2 - \Pi_{+}^2 - \Pi_{-}^2 - \frac{48}{11}\Pi_\phi^2 +
\frac{18}{11}\Pi_\phi\,\Pi_\sigma - \frac{1}{11}\Pi_\sigma^2 - 384\,\pi\,G_{N}\,C_{\gamma}\,e^{3(1-\gamma)\Omega}
- 12\,e^{4(\Omega + \beta_{+} + \sqrt{3}\beta_{-})}\right].
\label{ham}
\end{equation}
We can introduce a new set of variables that involving the Misner variables $(\Omega,\beta_{+},\beta_{-})$ in the
gravitational part \footnote[1]{Here $e^{\beta_1+\beta_2+\beta_3}=e^{3\Omega}=V$ corresponds to the volume of the Bianchi type
II Universe, in similar way that the flat Friedmann-Robetson-Walker metric (FRW) with the scale factor $A$.},

\begin{alignat}{2}
\beta_{1}&=\Omega + \beta_{+} + \sqrt{3}\beta_{-}, &\qquad \Omega&=\frac{1}{3}\left(\beta_{1} + \beta_{2} + \beta_{3}\right),\nonumber\\
\beta_{2}&=\Omega + \beta_{+} + \sqrt{3}\beta_{-}, &\qquad \beta_{+}&=\frac{1}{6}\left(\beta_{1} + \beta_{2} - 2\beta_{3}\right),\label{newvaror}\\
\beta_{3}&=\Omega - 2\beta_{+}, &\qquad \beta_{-}&=\frac{\sqrt{3}}{6}\left(\beta_{1} - \beta_{2}\right).\nonumber
\end{alignat}
with the last set of new variables \eqref{newvaror}, the Langrangian density \eqref{lagra-i} can be transformed as

\begin{equation}
\mathcal{L}_{II}=\frac{e^{\beta_1+\beta_2+\beta_3}}{N}\biggl(2\,\dot{\phi}^2
+ 36\,\dot{\phi}\,\dot{\sigma} + 96\,\dot{\sigma}^2
+ 2\,\dot{\beta}_{1}\,\dot{\beta}_{2} + 2\dot{\beta}_{1}\dot{\beta}_{3} + 2\dot{\beta}_{3}\,\dot{\beta}_{2}
+ 16\pi GN^2 C_{\gamma}\,e^{-(1 + \gamma)(\beta_{1} + \beta_{2} +
\beta_{3})}+\frac{1}{2} e^{2(\beta1-\beta_2-\beta_3)}\biggr).
\end{equation}
and the Hamiltonian density in the new variables is

\begin{align}
\mathcal{H}_{II}=\frac{1}{8}\,e^{-(\beta_{1} + \beta_{2} + \beta_{3})}\biggl[&-\Pi^2_{1} - \Pi^2_{2} - \Pi^2_{3} +
2\,\Pi_{1}\Pi_{2} + 2\,\Pi_{1}\Pi_{3} + 2\,\Pi_{2}\Pi_{3} -
\frac{16}{11}\,\Pi^2_{\phi}
+ \frac{6}{11}\,\Pi_{\phi}\,\Pi_{\sigma}
-\frac{1}{33}\,\Pi^2_{\sigma}\nonumber\\
&- 128\,\pi\,G\, C_{\gamma}\,e^{(1 - \gamma)(\beta_{1} + \beta_{2} +
\beta_{3})} \biggr] -\frac{1}{2}e^{-(\beta_{1} + \beta_{2} +
\beta_{3})}e^{4\beta_1}. \label{hami}
\end{align}
In the following sections we will analyze each one of the case that involve the terms in our classical Hamiltonian
\eqref{hami}. We will split our analysis in two subsections.

\section{Case of interest}
So far, we have built the Hamiltonian density of the Lagrangian \eqref{lagra-i} in terms of a new set of variables
\eqref{newvaror}, but we need to analyze the behavior of each one of the cases, with and without matter, these corresponds to
the vacuum case $(C_{\gamma}=0)$ and stiff fluid $(\gamma=1)$. We start our analysis with

\begin{enumerate}
\item \emph{Vacuum case: $C_\gamma=0$.}\\
We can see that the classical Hamiltonian can be written as

\begin{align}
\mathcal{H}_{IIvacuum}&=\frac{1}{8}e^{-(\beta_{1} + \beta_{2} + \beta_{3})}\left[2\Pi_{2}\Pi_{3} + 2\Pi_{1}\Pi_{2}
+ 2\Pi_{1}\Pi_{3} - \Pi_{1}^2 - \Pi_{2}^2 - \Pi_{3}^2 + \frac{6}{11}\Pi_{\phi}\Pi_{\sigma} - \frac{1}{33}\Pi_{\sigma}^2
- \frac{16}{11}\Pi_{\phi}^2\right]\nonumber\\
&-\frac{1}{2}e^{-(\beta_{1} + \beta_{2} + \beta_{3})}\,e^{4\beta_{1}}.
\label{htwovac}
\end{align}
By choosing time-gauge condition on the lapse function $N=e^{(\beta_{1}+\beta_{2}+\beta_{3})}$, the Hamilton's equation
associated to the Hamiltonian \eqref{htwovac} are given by

\begin{alignat}{3}
\label{pieqmo}\Pi_{1}'&=-2\,e^{4\beta_{1}}, &\quad \Pi_{2}'&=0, &\quad \Pi_{3}'&=0,\\
\Pi_{\phi}'&=0, &\quad \Pi_{\sigma}'&=0, &\quad \phi'&=\left(\frac{3}{44}\Pi_{\sigma} - \frac{4}{11}\Pi_{\phi}\right),\label{pisigma}\\
\label{momcomb}\sigma'&=\left(\frac{3}{44}\Pi_{\phi} - \frac{1}{132}\Pi_{\sigma}\right), &\quad \beta_{1}'&=\frac{1}{4}\left(-\Pi_{1} + \Pi_{2} + \Pi_{3}\right),
 &\quad
\beta_{2}'&=\frac{1}{4}\left(\Pi_{1} - \Pi_{2} + \Pi_{3}\right),\\
\label{mombethree}\beta_{3}'&=\frac{1}{4}\left(\Pi_{1} + \Pi_{2} - \Pi_{3}\right).
\end{alignat}
From the equations \eqref{pieqmo} we have that the momenta associated to $\beta_{2}$ and $\beta_{3}$ are given by

\begin{equation}
\Pi_{1}=a=const,  \qquad \Pi_{2}=b=const.
\label{solpige}
\end{equation}
Given the above restrictions \eqref{solpige}, we obtain a differential equation for $\Pi_{1}$

\begin{equation}
2\Pi'_{1} - \Pi_{1}^2 + \nu\Pi_{1} + \beta = 0,
\label{eqpi1}
\end{equation}
where $\nu=2(a+b)$, $\beta=\phi_{0} - (a-b)^2$, and $\phi_{0}=\frac{6}{11}\phi_{1}\sigma_{1} - \frac{1}{33}\sigma_{1}^2
- \frac{16}{11}\phi_{1}^2$. The solution to the differential equation \eqref{eqpi1} is

\begin{equation}
\Pi_{1}= \frac{1}{2}\nu - \frac{1}{2}\sqrt{\nu^2 + 4\beta}\,\tanh\left(-\frac{1}{4}\sqrt{\nu^2 + 4\beta}\,t + D\right)
\label{solpi1}
\end{equation}
where $D$ is an integration constant.
\\*
With this in mind, we see from the \eqref{momcomb}, and \eqref{mombethree} that

\begin{equation}
\frac{d}{dt}(\beta_{1}+\beta_{2}+\beta_{3})=\frac{1}{4}\nu - \frac{1}{2}\sqrt{\nu^2+4\beta}\tanh\left(
-\frac{1}{4}\sqrt{\nu^2+4\beta}\,t+D\right)
\label{soltosumbe}
\end{equation}
Integrating the last expression, the solution associated to $\beta_{1} + \beta_{2} + \beta_{3}$ is



\begin{equation}
\beta_{1} + \beta_{2} + \beta_{3}=\frac{1}{4}\nu\Delta t + \frac{1}{2}\frac{\sqrt{\nu^2+4\beta}}{\sqrt{\nu^2+4\beta}-4D}
\ln\cosh\left(-\frac{1}{4}\sqrt{\nu^2+4\beta}\,t+D\right).
\label{sumbetas}
\end{equation}
From the expressions \eqref{pieqmo}, \eqref{momcomb}, and \eqref{mombethree} we can see that the reparametrization of the
Misner variables \eqref{newvaror} are given by

\begin{eqnarray}
\beta_{1}&=&\frac{\sqrt{\nu^2+4\beta}}{8D - 2\sqrt{\nu^2 + 4\beta}}\ln\cosh\left(-\frac{1}{4}\sqrt{\nu^2+4\beta}\,t + D\right),\\
\beta_{2}&=&\frac{1}{2}b\Delta t - \frac{\sqrt{\nu^2+4\beta}}{8D - 2\sqrt{\nu^2 + 4\beta}}\ln\cosh\left(-\frac{1}{4}\sqrt{\nu^2+4\beta}\,t + D\right),\\
\beta_{3}&=&\frac{1}{2}a\Delta t - \frac{\sqrt{\nu^2+4\beta}}{8D - 2\sqrt{\nu^2 + 4\beta}}\ln\cosh\left(-\frac{1}{4}\sqrt{\nu^2+4\beta}\,t + D\right),\\
\Omega&=&\frac{1}{12}\nu\Delta t + \frac{1}{6}\frac{\sqrt{\nu^2 + 4\beta}}{\sqrt{\nu^2 + 4\beta} - 4D}\ln\cosh\left(-\frac{1}{4}\sqrt{\nu^2+4\beta}\,t + D\right).
\end{eqnarray}

\item \emph{Stiff fluid case: $\gamma=1$. }\\
In this case we have that the Hamiltonian is the expression \eqref{hami} with $\gamma=1$, so

\begin{align}
\mathcal{H}_{II}=\frac{1}{8}\,e^{-(\beta_{1} + \beta_{2} +
\beta_{3})}\biggl[&-\Pi^2_{1} - \Pi^2_{2} - \Pi^2_{3} +
2\,\Pi_{1}\Pi_{2} + 2\,\Pi_{1}\Pi_{3} + 2\,\Pi_{2}\Pi_{3} -
\frac{16}{11}\,\Pi^2_{\phi} + \frac{6}{11}\,\Pi_{\phi}\,\Pi_{\sigma}
-\frac{1}{33}\,\Pi^2_{\sigma}\nonumber\\
&- \rho_1 \biggr] -\frac{1}{2}e^{-(\beta_{1} + \beta_{2} +
\beta_{3})}e^{4\beta_1}, \qquad \rho_1=128\,\pi\,G\, C_1
\label{hami-1}
\end{align}
and the Hamilton equations are given by the same equations found in the before case \eqref{pieqmo}, \eqref{pisigma},
\eqref{momcomb} and \eqref{mombethree}, so the solutions become the same, only suffer change in the constant $\beta$ where is
necessary to put the constant $\rho_1$ for the content matter in the stiff scenario.
\end{enumerate}
The volume of the universe for this cosmological model become as

\begin{equation*}
V=e^{3\Omega}=\rm e^{\frac{1}{4}\nu t}\,\, \cosh\left[-\frac{1}{4}\sqrt{\nu^2+4\beta}\,t + D
\right]^{\frac{1}{2} \frac{\sqrt{\nu^2 + 4\beta}}{\sqrt{\nu^2 + 4\beta} - 4D}},
\end{equation*}
which has a growing trend for any time.

\section{Quantum Wheeler-DeWitt (WDW) formalism}
We find in the literature a lot of works on the Wheeler-DeWitt (WDW) equation
dealing with different problems, for example in \cite{gibbons}, they
asked the question of what a typical wave function for the universe
is. In Ref. \cite{ruffini} there appears an excellent summary of a
paper on quantum cosmology where the problem of how the universe
emerged from big bang singularity can no longer be neglected in the
GUT epoch. On the other hand, the best candidates for quantum
solutions become those that have a damping behavior with respect to
the scale factor, represented in our model with the $\Omega$
parameter, in the sense that we obtain a good classical solution
using the WKB approximation in any scenario in the evolution of our
universe \cite{Hartle,hawking}.

The WDW equation for this model is achieved by replacing the momenta $\Pi_{q^\mu}=-i \partial_{q^\mu}$, associated to the
Misner variables $(\Omega,\beta_{+}, \beta_{-})$ and the moduli fields $(\phi, \sigma)$ in the Hamiltonian \eqref{ham}. The
factor $e^{-3\Omega}$ may be factor ordered with $\hat\Pi_\Omega$ in many ways. Hartle and Hawking \cite{Hartle} have suggested
what might be called a semi-general factor ordering which in this case would order $e^{-3\Omega} \hat \Pi^2_\Omega$ as

\begin{equation}
- e^{-(3- Q)\Omega}\, \partial_\Omega e^{-Q\Omega}
\partial_\Omega= - e^{-3\Omega}\, \partial^2_\Omega + Q\,
e^{-3\Omega} \partial_\Omega\,\,,
\label {fo}
\end{equation}
where  Q is any real constants that measure the ambiguity in the factor ordering in the variable $\Omega$ and the corresponding
momenta. We will assume in the following this factor ordering for the Wheeler-DeWitt equation, which becomes
\begin{equation}
\Box \, \Psi + Q \frac{\partial \Psi}{\partial
\Omega}-\frac{48}{11}\frac{\partial^2 \Psi}{\partial \phi^2}
-\frac{1}{11}\frac{\partial^2 \Psi}{\partial
\sigma^2}+\frac{18}{11}\frac{\partial^2 \Psi}{\partial\phi \partial
\sigma}  -\left[b_\gamma e^{3(1-\gamma)\Omega} + 12\,e^{4(\Omega +
\beta_{+} + \sqrt{3}\beta_{-})}\right]\Psi =0, \label{WDW}
\end{equation}
where $\Box$ is the  three dimensional d'Lambertian in the $\ell^\mu=(\Omega,\beta_+,\beta_-)$ coordinates, with signature
$(- + +)$. On the other hand, we could interpreted the WDW equation \eqref{WDW} as a time-reparametrization invariance of the
wave function $\Psi$. At a glance, we can see that the WDW equation is static, this can be understood as the problem of time
in standard quantum cosmology. We can avoid this problem by measuring the physical time with respect a kind of time variable
anchored within the system, that means that we could understand the WDW equation as a correlation between the physical time
and a fictitious time \cite{bae,paulo-vargas}.\\*
When we introduce the ansatz $\Psi = \Phi(\phi,\sigma)\Theta(\Omega,\beta_\pm)$ in \eqref{WDW}, we obtain the general set of
differential equations (under the assumed factor ordering),

\begin{eqnarray}
\Box\Theta + Q \frac{\partial \Theta}{\partial \Omega} -
\left[b_\gamma e^{3(1-\gamma)\Omega}+ 12\,e^{4(\Omega + \beta_{+} +
\sqrt{3}\beta_{-})}-\mu^2\right]\,
\Theta &=& 0,\label{wdwmod}\\
\frac{48}{11}\,\frac{\partial^2 \Phi}{\partial \phi^2}
+\frac{1}{11}\,\frac{\partial^2 \Phi}{\partial
\sigma^2}-\frac{18}{11}\,\frac{\partial^2 \Phi}{\partial\phi
\partial \sigma} +\mu^2 \Phi&=&0. \label {phi-1}
\end{eqnarray}
The solution to the moduli fields corresponds to the hyperbolic partial differential equation \eqref{phi-1}, given by

\begin{equation}
\Phi(\phi,\sigma)= C_{1}\,\sin(C_{3}\,\phi + C_{4}\,\sigma + C_{5})
+ C_{2}\,\cos(C_{3}\,\phi + C_{4}\,\sigma + C_{5}) \label{solphi}
\end{equation}
where $\{C_i\}_{i=1}^5$ are integration constants and they are in
terms of $\mu$. We claim that this solutions is the same for all
Bianchi Class A cosmological models, because the Hamiltonian
operator in (\ref{WDW}) can be written  in separated way as $\rm
\hat H(\Omega,\beta_\pm,\phi,\sigma)\Psi=\hat
H_g(\Omega,\beta_\pm)\Psi + \hat H_m(\phi,\sigma)\Psi=0$, where $\rm
\hat H_g$ y $\rm \hat H_m$ represents the Hamiltonian to
gravitational sector and  the moduli fields, respectively.

We can rewrite the expression \eqref{wdwmod} in terms of the
Misner's variables, this means that we need to transform the
d'Lambertian operator $\Box$ in terms of the new variables
\eqref{newvaror}. Since the d'Lambertian operator in terms of the
Misner variables $(\Omega, \beta_{+}, \beta_{-})$ is given by

\begin{equation}
\Box=-\frac{\partial^2}{\partial\Omega^2} + \frac{\partial^2}{\partial\beta_{+}^2} + \frac{\partial^2}{\partial\beta_{-}^2}.
\label{dlam}
\end{equation}
So, we can transform the d'Lambertian operator \eqref{dlam} in terms of the new variables as

\begin{equation}
\Box:= 3\left(\frac{\partial^2}{\partial\beta_{1}^2} +
\frac{\partial^2}{\partial\beta_{2}^2} +
\frac{\partial^2}{\partial\beta_{3}^2}\right) -
6\left(\frac{\partial^2}{\partial\beta_{1}\partial\beta_{2}} +
\frac{\partial^2}{\partial\beta_{1}\partial\beta_{3}} +
\frac{\partial^2}{\partial\beta_{2}\partial\beta_{3}}\right)
\end{equation}
Finally, the WDW equation \eqref{wdwmod} is given by

\begin{align}
&&3\left[\frac{\partial^2\Theta}{\partial\beta_{1}^2} +
\frac{\partial^2\Theta}{\partial\beta_{2}^2}+
\frac{\partial^2\Theta}{\partial\beta_{3}^2}\right]+
Q\left(\frac{\partial\Theta}{\partial\beta_{1}}  +
\frac{\partial\Theta}{\partial\beta_{2}}  +
\frac{\partial\Theta}{\partial\beta_{3}}\right) -
6\left(\frac{\partial^2\Theta}{\partial\beta_{1}\partial\beta_{2}}
+ \frac{\partial^2\Theta}{\partial\beta_{1}\partial\beta_{3}} + \frac{\partial^2\Theta}{\partial\beta_{2}\partial\beta_{3}}\right)\nonumber\\
&&- \biggl[b_{\gamma}e^{(1-\gamma)(\beta_{1} + \beta_{2} +
\beta_{3})}  + 12 e^{4\beta_{1}}-\mu^2\biggr]\Theta=0.
\label{newWDW}
\end{align}

\subsection{Bianchi II with $\gamma=1$. }
With this consideration, the expression \eqref{newWDW} is reduced to

\begin{equation}
3\frac{\partial^2\Theta}{\partial\beta_{1}^2} +
Q\frac{\partial\Theta}{\partial\beta_{1}} +
3\frac{\partial^2\Theta}{\partial\beta_{2}^2} +
Q\frac{\partial\Theta}{\partial\beta_{2}} +
3\frac{\partial^2\Theta}{\partial\beta_{3}^2} +
Q\frac{\partial\Theta}{\partial\beta_{3}} -
6\biggl(\frac{\partial^2\Theta}{\partial\beta_{1}\partial\beta_{2}}
+ \frac{\partial^2\Theta}{\partial\beta_{1}\partial\beta_{3}} +
\frac{\partial^2\Theta}{\partial\beta_{2}\partial\beta_{3}}\biggr)
+\left(\mu_0^2 - 12 e^{4\beta_{1}}\right)\Theta=0, \label{eqthered}
\end{equation}
with $\mu_0^2=\mu^2-b_1$. The last partial differential equation
has a solution by taking the following ansatz

\begin{equation}
\Theta(\beta_{1},\beta_{2},\beta_{3})=e^{A\beta_{1}+B\beta_{2}+C\beta_{3}}\,G(\beta_{1}),
\label{ansthered}
\end{equation}
where $A, B, C$ are arbitrary constants and $G(\beta_{1})$ is an arbitrary functions depending on the variable $\beta_{1}$.
Substituting this ansatz into \eqref{eqthered}, we find that the $G$ function satisfy the ordinary differential equation

\begin{equation*}
\frac{d^2 G}{d\beta_1^2} + \alpha_0 \frac{dG}{d\beta_1}+\left[ \alpha -4 e^{4\beta_1}\right]G=0
\end{equation*}
with solution

\begin{equation}
\rm G(\beta_1)=
e^{-\frac{\alpha}{2}\beta_1}\,Z_\nu\left(e^{2\beta_1}\right).
\end{equation}
where $\rm Z_\nu(z)$ are the modified Bessel functions $\rm
(K_\nu(z),I_\nu(z))$,  $\rm \alpha_0=\frac{Q}{3}+2A-2(B+C)$, $\rm
\alpha=\frac{Q}{3}(A+B+C)+A^2+B^2+C^2-2(AB+AC+BC)-4\mu_0^2$,  and
$\rm \nu=\frac{1}{12}\sqrt{Q^2 -24QB - 24QC + 144BC-4\mu_0^2}$ its
order. In order that the unnormalized probability density
$|\Theta(\Omega,\beta_\pm)|^2$ does not diverge for $\rm
|\beta_\pm|\to \infty$ and at fixed $\Omega$, which are the
gravitational variables, we drop the function $\rm I_\mu(z)$, remain
only the function $\rm K_\mu(z)$.  So, we have that the solution to
the partial differential equation (PDE) \eqref{eqthered} is given by

\begin{equation}
\rm \Theta(\beta_{1} + \beta_{2} + \beta_{3})=e^{\frac{1}{6}(6B + 6C
- Q+12\mu_0^2)\beta_{1}+B\beta_2 + C\beta_3}\,K_\nu
\left(e^{2\beta_{1}}\right), \label{partsol}
\end{equation}
The asymptotic solution for large argument to this Bessel function, goes as

\begin{equation}
\rm \Theta(\beta_1)\approx e^{\frac{1}{6}(6B + 6C -
Q+12\mu_o^2)\beta_{1}+B\beta_2 + C\beta_3}\, e^{-e^{2\beta_1}}.
\label{asymptotic}
\end{equation}
This solution we will be compared with solution obtained using the
Bohm's formalism \cite{bohm}. Employing this formalism we find that
its amplitude of probability $\rm W$ is given by

\begin{equation}
\rm W=e^{\frac{1}{6}(6B + 6C - Q+12\mu_0^2)\beta_{1} + B\beta_{2} +
C\beta_{3}}. \label{amproII}
\end{equation}
For the particular case, where \eqref{eqthered} just is depending on the variable $\beta_{1}$ we have that its solution is

\begin{equation}
\Theta=e^{(-\frac{1}{6}Q+2\mu_0^2)\beta_{1}}\,K_{\frac{\sqrt{Q^2-4\mu_0^2}}{12}}(e^{2\beta_{1}}).
\label{solob}
\end{equation}

\subsection{Solution in the Bohm's formalism when $\mu_0^2=0$}
We present the main ideas of this formalism to solve the WDW
equation, you can see \cite{intech}. Also we use the hidden symmetry
in the potential U for this model \cite{graham,sukumar}, which seems
to be a general property of the Bianchi models.

 We use the following
ansatz for the wave function
\begin{equation}
\Theta(\ell^\mu) = W(\ell^\mu) e^{- S(\ell^\mu)},
\label{ans}
\end{equation}
where $S(\ell^\mu)$ is known as the superpotential function, and W is the amplitude of probability which is used in Bohmian
formalism \cite{bohm}, those found in the literature, years ago \cite{os}. So \eqref{wdwmod} is transformed into

\begin{equation}
\Box\,W - W\Box\,S - 2\nabla W\cdot\nabla S +
Q\,\frac{\partial W}{\partial\Omega} - Q\,W\,\frac{\partial S}{\partial\Omega} +
W\left[\left(\nabla S\right)^2 - U\right] = 0,
\label{mod}
\end{equation}
where  $\rm \Box = G^{\mu \nu}\frac{\partial^2}{\partial \ell^\mu\partial \ell^\nu}$,
$\rm {\nabla \, W}\cdot {\nabla \, \Phi}=G^{\mu\nu} \frac{\partial W}{\partial \ell^\mu}\frac{\partial\Phi}{\partial \ell^\nu}$
, $\rm (\nabla)^2= G^{\mu\nu}\frac{\partial }{\partial \ell^\mu}\frac{\partial }{\partial
\ell^\nu}= -(\frac{\partial}{\partial \Omega})^2
+(\frac{\partial}{\partial \beta_+})^2 + (\frac{\partial}{\partial
\beta_-})^2,$ with $\rm G^{\mu \nu}= diag(-1,1,1)$,  $U$ is the potential term of the cosmological model under consideration.
Eq. \eqref{mod} can be written as the following set of partial differential equations

\begin{subequations}
\label{WDWa}
\begin{eqnarray}
(\nabla S)^2 - U &=& 0, \label{hj} \\
  W \left( \Box S + Q \frac{\partial S}{\partial \Omega}
  \right) + 2 \nabla \, W \cdot \nabla \, S &=& 0 \, ,\label{wdwho} \\
  \Box \, W + Q \frac{\partial W}{\partial \Omega} & = & 0. \label{cons}
\end{eqnarray}
\end{subequations}
Following reference \cite{wssa}, first we shall choose to solve Eqs. \eqref{hj} and \eqref{wdwho}, whose solutions at the end
will have to fulfill Eq. \eqref{cons}, which play the role of a constraint equation.

\subsection{Transformation of the Wheeler-DeWitt equation}
We were able to solve \eqref{hj}, by doing the change of coordinates \eqref{newvaror} and rewrite \eqref{hj} in these new
coordinates, with this change, the function S is obtained by taking the ansatz \eqref{ans}.In this section, we shall
obtain the solutions to the equations that appear in the decomposition of the WDW equation, \eqref{hj},
\eqref{wdwho} and \eqref{cons}, using the Bianchi type II Cosmological model. So, the equation $[\nabla]^2= -(\frac{\partial}{\partial
\Omega})^2 +(\frac{\partial}{\partial \beta_+})^2 +(\frac{\partial}{\partial \beta_-})^2$ can be written in the following way

\begin{eqnarray}
\left[\nabla \right]^2 &=& 3 \left[ \left(\frac{\partial}{\partial
\beta_1}\right)^2+ \left(\frac{\partial}{\partial \beta_2}
\right)^2+ \left(\frac{\partial}{\partial \beta_3}\right)^2 \right ]
- 6\left[
 \frac{\partial}{\partial \beta_1} \frac{\partial}{\partial \beta_2} +
 \frac{\partial}{\partial \beta_1} \frac{\partial}{\partial \beta_3} +
 \frac{\partial}{\partial \beta_2} \frac{\partial}{\partial \beta_3}
\right ]\nonumber\\
 &=& 3  \left ( \frac{\partial}{\partial \beta_1}+
                   \frac{\partial}{\partial \beta_2}+
                   \frac{\partial}{\partial \beta_3} \right )^2 -12 \left [
\frac{\partial}{\partial \beta_1} \frac{\partial}{\partial \beta_2}
+ \frac{\partial}{\partial \beta_1} \frac{\partial}{\partial
\beta_3} + \frac{\partial}{\partial \beta_2}
\frac{\partial}{\partial \beta_3} \right ]. \label {nab}
\end{eqnarray}
The potential term of the Bianchi type II is transformed in the new variables into

\begin{equation}
U = 12\,e^{4\beta_1}.
\label{pot}
\end{equation}
Then \eqref{hj} for this models is rewritten  in the new variables as

\begin{equation}
3\left(\frac{\partial S}{\partial \beta_1} + \frac{\partial S}{\partial \beta_2} + \frac{\partial S}{\partial \beta_3} \right)^2 -12 \left[
\frac{\partial S}{\partial \beta_1} \frac{\partial S}{\partial\beta_2} + \frac{\partial S}{\partial \beta_1} \frac{\partial
S}{\partial \beta_3} + \frac{\partial S}{\partial \beta_2} \frac{\partial S}{\partial \beta_3}\right] - 12  e^{4\beta_1} =0.
\label{hanv}
\end{equation}
Now, we can use the separation of variables method to get solutions to the last equation for the $S$ function, obtaining for
the Bianchi type II model

\begin{equation}
S_{II}= \pm  2e^{2\beta_1}.
\label{phi2}
\end{equation}
With this result, and using for the solution to \eqref{wdwho} in the new coordinates $\beta_i$, we have  for W function as

\begin{equation}
W_{_{II}}= W_0 \, e^{(1+ \frac{Q}{2})\beta_1},
\label{w2}
\end{equation}
and  re-introducing this result into  Eq. \eqref{cons} we find the constriction on the Q factor ordering, $Q=-2$ and
$Q=-\frac{2}{3}$, thus the equation \eqref{w2} have the form

\begin{equation}
W_{_{II}}=\left\{ \begin{tabular}{ll} $W_0$, &\qquad for
$Q=-2$\\
$W_0\, e^{ \frac{2}{3}\beta_1}$,& \qquad for $Q=-\frac{2}{3}$
\end{tabular}
\right.
\label{w2extra}
\end{equation}
So, the Bohm's solutions becomes (we demanding that $\Theta$ does
not diverge for $(\rm |\beta_+|, |\beta_-|)\to \infty$, for fixed
$\Omega$, the contribution $\rm e^{+ e^{2\beta_1}}$ is dropped)

\begin{equation}
\Theta(\beta_1)=W_{II}\,exp[-e^{2\beta_1}].
\label{w2extra1}
\end{equation}
This class of solution become as the pure bosonic first term and the
simple fermionic part in the decomposition of the wave function in
the supersymmetric quantum cosmology in the Grassmann
representation, for this same cosmological model, studied 22 years
ago \cite{osb}.

\begin{figure}[ht!]
\begin {center}
\includegraphics[totalheight=0.75\textheight]{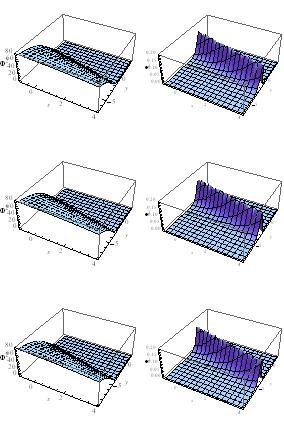}
\caption{ Full anisotropy (left)  and the lowest-order WKB
approximation (right) in the probability density corresponding to
the wave functions  to equation (\ref{solob}) and (\ref{w2extra1}),
respectively, considering the anisotropic variables $\rm
(x,y)\to(\beta_+,\beta_-)$, centering in $\Omega=-2,0$ and 2,
beginning in the up to down in the graphics. The numerical order
$\frac{1}{18}$ is considered for the value $Q=-\frac{2}{3}$ and the
constant $\rm \mu^2=0$ appeared in the Bohm's formalism.}
\label{fig}
\end{center}
\end{figure}

When we compare this solution with the general solution found using
separation of variables, we obtain that this corresponds to the
asymptotic behavior of this solution, that in some way corresponds
to the lowest-order WKB approximation \cite{sm,bae}, which is shown
in the Figure \ref{fig}, where the picks of the plot follow the
classical trajectory $\beta_+=-\sqrt{3}\beta_{-}-|\Omega|$, which
corresponds at zero phase in the probability amplitude in both
solutions; in other words, this wave function \eqref{w2extra1} has
translational symmetry along the last trajectory mentioned. Also is
presented a shifting to the wave function in the $\beta_-$ axe in
the $-\infty$ direction, how is shown in the graphics to the
probability density $\rm |\Theta|^2$ in both sectors (full
anisotropy and lowest-order WKB approximation).

With the conditions mentioned above, we shall calculate $|\Theta|^2$ as an unnormalized probability density, considering only
the anisotropy variables and the $\Omega$ parameter fixed. In the
full wave function we have an unnormalized behavior, however in the
Bohm's sector we have that the maximum is minor to unity, due that
only we consider the lowest-order WKB approximation.

\section{Final Remarks}
In this work we have explored a compactification of a
ten-dimensional gravity theory coupled with a time-dependent dilaton
into a time-dependent six-dimensional torus. The effective theory
which emerges through this process resembles in the Einstein frame
was applied to anisotropic cosmological  Bianchi type II model. By
incorporating the matter content and by using the analytical
procedure of Hamilton equation of classical mechanics, in
appropriate coordinates, we found the classical solution for the
anisotropic Bianchi type II cosmological model.  In the quantum
formalism for the standard Wheeler-DeWitt equation, we can observe
that this anisotropic model is completely integrable without employ
numerical methods, similar solutions to partial differential
equation have been found in \cite{SOCORRO2010}. In order to have the
best candidates for quantum solutions become those that have a
damping behavior with respect to the scale factor, represented in
our model with the $\Omega$ parameter, in this way we dropped in the
full solution the modified Bessel function $\rm I_\nu(z)$ and the
term $\rm e^S$ in the Bohm's solution. We obtain the relation
between these both  physical solutions in the asymptotic behavior to
the modified Bessel function for large argument, \eqref{asymptotic}
and it is compared with the Bohm's solutions, \eqref{w2extra1}, who
us indicate that is the lowest-order WKB approximation how is
indicate in the literature \cite{sm,bae}. The unnormalized
probability density which is shown in the Figure \ref{fig}, show
that the picks of the plot follow the classical trajectory
$\beta_+=-\sqrt{3}\beta_{-}-|\Omega|$, which corresponds at zero
phase in the probability amplitude in both solutions; in other
words, this wave function \eqref{w2extra1} has translational
symmetry along the last trajectory mentioned. Also is presented a
shifting to the wave function in the $\beta_-$ axe in the $-\infty$
direction, that in some sense us indicate that the anisotropy of the
model in the quantum state finish for large $+\beta_-$ and follow
this in the line $\beta_+=-\sqrt{3}\beta_{-}-|\Omega|$ .

For future work is to consider a more complete compactification
process in which all moduli are considered as time-dependent fields
as well as time-dependent fluxes and follow the analysis presented
in the ref. \cite{bae} to obtain the other orders to the WKB
approximation, and see how these orders contribute to the full
solution found by separation of variables.

\acknowledgments{ \noindent We thank to O. Loaiza for useful
comments to this work. This work was partially supported by CONACYT
167335, 179881 grants. PROMEP grants UGTO-CA-3. This work is part of
the collaboration within the Instituto Avanzado de Cosmolog\'{\i}a
and Red PROMEP: Gravitation and Mathematical Physics under project
\emph{Quantum aspects of gravity in cosmological models,
phenomenology and geometry of space-time}. One of authors (LTS) was
supported by a PhD scholarship in the graduate program by CONACYT.
Many calculations where done by Symbolic Program REDUCE 3.8, and
Cadabra \cite{Peeters,KaPe} into Maple software.}


\end{document}